\documentclass{article}

\usepackage{graphicx}
\usepackage{natbib}
\usepackage{amsmath}
\usepackage{amsfonts}

\newcommand{\R}{\mathbb R}
\newcommand{\N}{\mathbb N}
\newcommand{\n}{\underline{n}}

\newcommand{\norm}[1]{\left\Vert#1\right\Vert}
\newcommand{\abs}[1]{\left\vert#1\right\vert}

\newcommand{\eps}{\varepsilon}

\newtheorem{definition}{Definition}

\begin{document}

\markboth{Jan Lorenz}
{Continuous Opinion Dynamics under Bounded Confidence: A Survey}

\title{Continuous Opinion Dynamics under Bounded Confidence: A Survey}

\author{Jan Lorenz\footnote{Department of Mathematics and Computer
    Science, University of Bremen, Bibliothekstra\ss{}e, 28329 Bremen,
    Germany, math@janlo.de; Present adress: ETH Z\"urich, Chair of
    Systems Design, Keuzplatz 5, 8032 Z\"urich, Switzerland, jalorenz@ethz.ch}}

\date{July 12, 2007}

\maketitle

\begin{abstract}
Models of continuous opinion dynamics under bounded confidence have been presented independently by Krause and Hegselmann and by Deffuant et al in 2000. They have raised a fair amount of attention in the communities of social simulation, sociophysics and complexity science. The researchers working on it come from disciplines as physics, mathematics, computer science, social psychology and philosophy. 

In these models agents hold continuous opinions which they can gradually adjust if they hear the opinions of others. The idea of bounded confidence is that agents only interact if they are close in opinion to each other. 
Usually, the models are analyzed with agent-based simulations in a
Monte-Carlo style, but they can also be reformulated on the agent's
density in the opinion space in a master-equation style. The contribution
of this survey is fourfold. First, it will
present the agent-based and density-based modeling frameworks including
the cases of multidimensional opinions and heterogeneous bounds of
confidence. Second, it will give the bifurcation diagrams of cluster
configuration in the homogeneous model with uniformly distributed initial
opinions. Third, it will review the several extensions and the evolving
phenomena which have been studied so far, and fourth it will state some open questions.

\end{abstract}

\centerline{opinion dynamics, continuous opinions, cluster formation,
  bifurcation patterns}
\centerline{PACS Nos.: 89.20.-a; 89.65.-s}

\section{Introduction}

The term ``opinion dynamics'' nowadays summarizes a wide class of
different models differing in heuristics, formalization as well as in the
phenomena of interest. The latter range from emergence of fads, minority opinion spreading, collective decision making, finding and not finding of consensus, emergence of political parties, minority opinion survival, emergence of extremism and so on.

This paper deals with these phenomena in models of continuous opinion
dynamics under bounded confidence. `Continuous' refers to the opinion
issue and not to the time. Thus opinions in continuous opinion dynamics
should be expressable in real numbers where compromising in the middle is
always possible. Example issues are prices, tax rates or predictions about macroeconomic variables. The political spectrum is also often mapped to a continuum from left to right wing.

In opinion dynamics one considers a set of agents where each holds an
opinion from a certain opinion space. She may change her opinion when she
gets aware of the opinions of others. In the physics literature discrete
opinion spaces (classically binary opinions)
\cite{Galam2002,Schweitzer2000,Sznajd-Weron2000} have dominated research
due to their striking analogy with spin systems. Sometimes they have been
extended to more than two spin values, which are ordered and thus get closer to continuous opinion dynamics \cite{Stauffer2002a,Stauffer2002}. 

In recent years two models of genuinely continuous opinion dynamics under
bounded confidence have raised the interest of the sociophysics
community: the models of Hegselmann and Krause
\cite{KrauseEtal1997,Krause2000,Hegselmann2002} and Deffuant, Weisbuch and others
\cite{Deffuant2000,Weisbuch2002}. Due to physicist's research some
progress in understanding the dynamics of these models has been made,
especially by introducing the master equation on the agents density in
the opinion space for these type of models. For a recent review in this broader context of opinion dynamics including discrete opinions from a physicist's perspective see \cite{StaufferEtal2006}.

Let us consider a population of agents which hold diverse opinions about
certain issues expressible in real numbers. Each agent is willing to
change her opinion if she hears the opinions of others by adjusting
towards those opinions. Every adjustment in terms of averaging is
possible due to the continuous nature of the opinions. Further on,
consider agents to have bounded confidence. That means an agent is only
willing to take those opinions into account, which differ less than a certain bound of confidence $\eps$ from her own opinion. 

The Deffuant-Weisbuch (DW) model and the Hegselmann-Krause (HK) model
both rely on the idea of repeated averaging under bounded
confidence. They differ in their communication regime. In the DW
model agents meet in random pairwise encounters after which they
compromise or not. In the HK model, each agent moves to the average
opinion of all agents which lie in her area of confidence (including
herself). Actually, the DW model contains another parameter which
controlls how close an agent moves to the opinion of the other. But it
has turned out that this parameter has only an effect on convergence time in the basic model. Therefore we neglect it in the basic analysis and discuss its impact afterwards.

The DW model was partly inspired by the famous Axelrod model about the dissemination of culture \cite{Axelrod1997} where something similar to the bounded confidence assumption is implemented. It has been developed in a project about improving agri-environmental policies in the European union.

The HK model has been presented by Krause
\cite{KrauseEtal1997,Krause2000} in a mathematical context as a nonlinear
version of older consensus models
\cite{DeGroot1974,Chatterjee1977,Lehrer1981}. It was analyzed through
computer simulations by Hegselmann and Krause \cite{Hegselmann2002} in
the context of social simulation and has gained a lot of attention since. 

The next section presents the two models in their original agent-based versions and in their density-based formulation inspired by statistical physics. Section 3 shows and explains the bifurcation diagrams of both models which serve as a reference for the review of several extensions in Section 4.

\section{The models}

From the heuristic description of the models one can either define
agent-based dynamics for a finite population of $n$ agents or
density-based dynamics for a density function which determines the agents
density in the opinion space. The latter approach can then be interpreted
as taking
the infinite limit $n\to\infty$ of agent-based models. It is a classical tool of statistical physics often called the derivation of a master equation or rate equation which was first applied to social systems bei Weidlich \cite{Weidlich1971,Weidlich2000}. 

Nevertheless, the basic ingredient of both approaches is a continuous
opinion space $S\subset\R^d$. The $d$ real numbers represent opinions on $d$ different subjects. Usually, only compact and convex sets are regarded as appropriate opinion spaces. The simplest case that we will
discuss is the one-dimensional interval $[0,1]$. But definitions
extend naturally to more dimensions. 

In an agent-based model the state variable is an opinion profile $x(t)
\in S^n$, which is a vector of vectors. For agent $i$, it contains the opinion vector $x^i\in S$. For an initial opinion profile $x(0)$ dynamics are defined recursively as $x(t+1) = f(t,x(t))$. The pair $(S,f)$ is then a discrete dynamical system. 

In a density based model, the state variable is a density function on the
opinion space $P(t,\cdot): S \to \R_{\geq 0}$ with $\int_S P(t,x)dx = 1$
for all $t$. Dynamics are then defined for a given initial density
function $P(0,\cdot)$ as the evolution of the density function in
time. Time is sometimes regarded as discrete and sometimes as
continuous. In the continuous-time case we consider the differential
equation $\frac{\partial}{\partial t}P(t,x) = g(P(t,\cdot))$ where $g$
operates on the space of density functions. Then we try to find solutions
of the differential equation analytically or with numerical solvers. In
the discrete-time case we replace the differential operator by the
difference operator and write $\Delta P(t+1,x) = P(t+1,x) - P(t,x) =
g(P(t,\cdot))$. We can then directly compute the solution recursively
$P(t+1,x) = P(t,x) + \Delta P(t+1)$. Actually, both discrete
\cite{Lorenz2005a,LorenzPhD2007} and continuous
\cite{Ben-Naim2003,Fortunato2005b} time were used to compute solutions
numerically. It turned out that discrete and continuous solutions of the
DW model are quite similar but discrete and continuous solutions of the
HK model differ qualitatively, which we will point out later. 

In the following we will define the agent-based DW and HK model in
detail, as well as their density-based counterparts. The definitions will
already include the extensions of multidimensional opinions and
heterogeneous bounds of confidence. In a first step it is enough to think
of the opinion space as the interval $[0,1]$ and to regard all bounds of
confidence as equal. 

\subsection{The Deffuant-Weisbuch model}

We proceed with the definition of the agent-based version of the DW model. 

\begin{definition} \textbf{(Agent-based DW model)}
\label{mod:DW}
Let there be $n\in\N$ agents and an appropriate opinion space $S\subset\R^d$. 
Given an initial profile $x(0) \in S^n$, bounds of confidence
$\eps_1,\dots, \eps_n > 0$, and a norm $\norm{\cdot}$ we define the \emph{agent-based DW process} as the random process $(x(t))_{t\in\N}$ that chooses in each time step $t\in\N$ two
random agents $i,j$ which perform the action
\begin{eqnarray*}
x^i(t+1) &=& \left\{
\begin{array}{ll}
    (x^j(t)+x^i(t))/2 & \hbox{if $\norm{x^i(t)-x^j(t)}\leq\eps_i$} \\
    x^i(t) & \hbox{otherwise.}
\end{array}
\right.\\
\end{eqnarray*}
The same for $x^j(t+1)$ with $i$ and $j$ interchanged.

If $\eps_1=\dots=\eps_n$ we call the model \emph{homogeneous}, otherwise \emph{heterogeneous}.
\end{definition}

It has been shown \cite{Lorenz2005} that the homogeneous process always
converges to a limit opinion profile (usually not in finite time). The
same is observed in simulations of the heterogeneous case, but a proof is
lacking. A limit profile in the homogeneous case is an opinion profile
where for each two opinions $x_i,x_j$ it holds that they are either equal
(belong to the same cluster) or have a distance larger than $\eps$. Thus
further changes are not possible regardless of the choice of $i$ and
$j$. Further on, it is easy to see that the average opinion over all agents is conserved during dynamics \cite{Urbig2004}, but only in the homogeneous case.

For example trajectories of the process see \cite{Deffuant2000,LorenzPhD2007}. 

\medskip

For the density-based DW model we extend the model defined in \cite{Ben-Naim2003} to populations of agents with heterogeneous bounds of confidence. 

\begin{definition} \textbf{(Density-based DW model)}
Let $S\subset \R^d$ be an appropriate opinion space, $[\eps_1,\eps_2]$ be
an interval of possible bounds of confidence and the initial density
function on the opinion space times the interval of bounds of confidence
be $P(0,\cdot,\cdot): S \times [\eps_1,\eps_2] \to [0,\infty]$. with
$\int_S\int_{\eps_1}^{\eps_2} dx d\eps P(0,x,\eps) =
1$. \footnote{$P(0,\cdot,\cdot)$ is a density function at time zero over
  the opinion space and the interval of bounds of confidence,
  $P(0,x,\eps)dxd\eps$ represents the proportion of agents which hold
  opinions in $[x,x+dx]$ and bounds of confidence in $[\eps,\eps+d\eps]$.} For
abbreviation we define the aggregated density as $P(t,x) =
\int_{\eps_1}^{\eps_2}d\eps P(t,x,\eps) $. With this we define the differential equation
\begin{align}\nonumber
\frac{\partial}{\partial t}& P(t,x,\eps) =  \int\limits_S dx_1 \left[ \int\limits_{\eps_1}^{\eps_2} d\bar\eps \right. \left(\frac{}{}\right.\int\limits_{\norm{x_1-x_2}\leq\bar\eps} dx_2  \left[\frac{}{}P(t,x_1,\bar\eps) P(t,x_2) \quad  + \right. \\ \nonumber
 & \hspace{6cm} \left.\left. P(t,x_1) P(t,x_2,\bar\eps)\frac{}{}\right] \delta(x-\frac{x_1-x_2}{2}) \right) \\
 & \left. - \int\limits_{\norm{x_1-x_2}\leq \eps} dx_2 \left( P(t,x_1,\eps)P(t,x_2)\delta(x - x_1) + P(t,x_1)P(t,x_2,\eps)\delta(x-x_2) \right)\right] \label{eq:DWmaster1}
\end{align}
The continuous-time density-based DW process is then defined as the solution of the initial value problem.
The discrete-time density-based DW process is the sequence $(P(t,\cdot,\cdot))_{t\in\N}$ recursively defined by 
\[
P(t+1,x,\eps) = P(t,x,\eps) + \Delta P(t,x,\eps)
\] 
with $\Delta P(t,x,\eps) = \frac{\partial}{\partial t} P(t,x,\eps)$.
\end{definition}

Notice that it is important to distinguish the aggregated density $P(t,x)$ defined above and the density for a given bound of confidence $P(t,x,\eps)$ to understand how an agent looks at all agents regardless of their bound of confidence and adjusts to the ones within her bound of confidence. 

For a homogeneous bound of confidence $\eps = [\eps_1,\eps_2]$ the dynamical Equation \eqref{eq:DWmaster1} reduces to
\begin{align}
\frac{\partial}{\partial t} P(t,x) =  \int\limits_S dx_1 \int\limits_{\norm{x_1-x_2}\leq\eps} dx_2 \left(\frac{}{}\right.P(t,x_1)P(t,x_2)&\left(\frac{}{}\right.\underbrace{2\delta(x-\frac{x_1+x_2}{2})}_{\textrm{fraction joining state $x$}} \nonumber \\
& -\underbrace{(\delta(x-x_1) + \delta(x-x_2))}_{\textrm{fraction
leaving state $x$}}\left.\frac{}{}\right)\left.\frac{}{}\right) \label{eq:DWmaster2}
\end{align}
which has been reported first in \cite{Ben-Naim2003} and gives a good starting point for understanding populations with heterogeneous bounds of confidence. The $\delta$-functions represent the position where the mass of the two opinions $x_1$ and $x_2$ jumps to.

Equation \eqref{eq:DWmaster2} can be transformed to a form free of $\delta$-distributions,
\begin{align}
\frac{\partial}{\partial t} P(t,x) =  \int\limits_{\norm{x-y}\leq\frac{\eps}{2}} dy\, 4P(t,y)P(t,2x-y) \quad - \quad \int\limits_{\norm{x-y}\leq\eps} dy\, 2P(t,y). \label{eq:DWmaster3}
\end{align}
One insight is that the gain term for opinion $x$ is a `bounded convolution' of $P(t,\cdot)$ with itself at the point $2x$. 

It is pointed out in \cite{Ben-Naim2003} that Equation
\eqref{eq:DWmaster2} conserves the mass and the mean opinion of the
density. For Equation \eqref{eq:DWmaster1} this holds only for the mass,
which is conserved for each value of $\eps$ and of course for the
aggregated density, too. This gives rise to the fact that under
heterogeneous bounds of confidence overall drifts (in terms of a drift of
the mean opinion) to more extremal opinions may happen. This has been simulated by \cite{Deffuant2002} in a very stylized model.

In \cite{Ben-Naim2003} it is shown for $S=[0,1]$ and $\eps\geq 0.5$ that
$P(t,\cdot)$ in Equation \eqref{eq:DWmaster2} goes with $t\to\infty$ to
a limit distribution $P(\infty,x) = \delta(x-x_0)$ where $x_0$ is the
mean opinion. For lower epsilon it is postulated (and verified by
simulation) that $P(\infty,x) = \sum_{i=1}^r m_i\delta(x-x_i)$ where $r$
is the number of evolving opinion clusters, $x_i\in [0,1]$ is the
position of cluster $i$ and $m_i$ is its mass. All clusters must fulfill
the conservation laws $\sum_{i=1}^r m_i = 1$ and $\sum_{i=1}^r x_im_i$ is
equal to the conserved mean opinion. Further on, all different clusters $i\neq j$ must fulfill $|x_i-x_j|>\eps$. This conicides with the result for the agent-based model.

In general the right hand side of Equation \eqref{eq:DWmaster2} does not depend continuously on $P(t,\cdot)$. \footnote{Consider $P(t,\cdot)$ = $\frac{1}{2}\delta(x-1) + \frac{1}{2}\delta(x+1+\xi)$ then we have a fixed point for $\eps=2$ and any $\xi>0$. But for $\xi = 0$ applying the right hand side of Equation \eqref{eq:DWmaster2} would give $\frac{1}{4}\delta(x-1) + \frac{1}{2}\delta(x) + \frac{1}{2}\delta(x+1)$.} But this happens only in the case where $P(t,\cdot)$ contains $\delta$-functions and is thus not a `normal' function in $x$ but a distribution. Noncontinuity of the right hand side of \eqref{eq:DWmaster2} implies that there is no or at least no unique solution. Fortunately, it turned out that $\delta$-functions do not evolve during the process but only in the limit $t\to\infty$. If we start with $P(t,\cdot)$ without $\delta$-functions trajectories can be computed as it is done in \cite{Ben-Naim2003} with a fourth order Adams-Bashforth algorithm. 

Moreover, it turned out in comparing results of \cite{Ben-Naim2003} and
\cite{LorenzPhD2007} that the discrete-time density-based process leads
to nearly the same limit densities as the continuous-time density-based
process. In \cite{Lorenz2005a,LorenzPhD2007}, additional to discrete
time, the space $P(t,\cdot)$ has been discretized to $p(t) \in \R^n_{\geq
  0}$. Dynamics can then be defined as an interactive Markov chain (a
state and time discrete Markov chain where transition probabilities
depend on the actual state). Essentially, the interactive Markov chain is the same what evolves by discretization of the continuous opinion space for numerical computation in \cite{Ben-Naim2003}. 

In a nutshell, there are different approaches but results for the density-based DW model lie all close together. This does not hold for the HK model as we will see. 

Examples for trajectories of the process can be found in \cite{LorenzPhD2007,Ben-Naim2003}.

\subsection{The Hegselmann-Krause model}
Here we will give the HK model, first in its original agent-based version.  

\begin{definition}\textbf{Agent-based HK model}\label{mod:HK}
Let there be $n\in\N$ agents and an appropriate opinion space $S\subset\R^d$. 

Given an initial profile $x(0) \in S^n$, bounds of confidence
$\eps_1,\dots, \eps_n > 0$ and a norm $\norm{\cdot}$ we define the
\emph{HK process} $(x(t))_{t\in\N}$ recursively
through
\begin{equation}
x(t+1) = A(x(t),\eps_1,\dots,\eps_n) x(t),
\end{equation}
with $A(x,\eps_1,\dots,\eps_n)$ being the \emph{confidence matrix} defined
\[
 A_{ij}(x,\eps_1,\dots,\eps_n) :=
 \left\{ \begin{array}{cl}
   \frac{1}{\#I_{\eps_i}(i,x)} \quad & \textrm{if } j\in I_{\eps_i}(i,x)   \\
   0 & \textrm{otherwise,}
\end{array} \right. \\
\]
with $I_{\eps_i}(i,x) := \{j\in\n \,|\, \norm{x^i - x^j} \leq \eps_i \}$ being the \emph{confidence set} of agent $i$ with respect to opinion profile $x$. In other words, the new opinion $x^i(t+1)$ is the arithmetic average over all opinions in $x(t)$ differing from it by not more than $\eps_i$.

If $\eps_1=\dots=\eps_n$ we call the model \emph{homogeneous}, otherwise \emph{heterogeneous}\index{heterogeneous repeated meeting process}.
\end{definition}

It has been shown \cite{Dittmer2001,Lorenz2003b} that the homogeneous
process always converges to a limit opinion profile in finite time. The
same convergence is observed in simulations for the heterogeneous case but
it does not occur in finite time and a proof is lacking. A limit profile in the
homogeneous case is always a fixed point $x^\ast =
A(x^\ast,\eps)x^\ast$. In $x^\ast$ it holds for each pair of two opinions
$x^i,x^j$ that they are either equal (belong to the same cluster) or have
a distance larger than $\eps$. The proof for this is not difficult but
also not trivial and can be found in \cite{LorenzPhD2007}. In
contrast to the DW model the mean opinion is not conserved. The mean opinion is only conserved when the initial profile is symmetric around its mean opinion. 

For example trajectories of the process see \cite{Hegselmann2002,LorenzPhD2007}.

We now turn our attention to the density-based version of the HK model for which we extend models independently developed in \cite{Fortunato2005b} and \cite{Lorenz2005a}. 

For the definition of the density-based HK model we need the definition of the \emph{$\eps$-local mean}
\[
M_1(x,P(t,\cdot),\eps) = \frac{\int_{x-\eps}^{x+\eps} yP(t,y) dy}{\int_{x-\eps}^{x+\eps} P(t,y) dy}.
\]
It gives the expected value of $P(t,\cdot)$ in the interval $[x-\eps,x+\eps]$. The nominator gives the first moment in the $\eps$-interval around $x$ while the denominator gives the necessary renormalization by the probability mass in that interval.

\begin{definition} \textbf{(Density-based HK model)}
Let $S\subset \R^d$ be an appropriate opinion space, $[\eps_1,\eps_2]$ be
an interval of possible bounds of confidence and the initial density
function on the opinion space times the interval of bounds of confidence
be $P(0,\cdot,\cdot): S \times [\eps_1,\eps_2] \to [0,\infty]$ with
$\int_S\int_{\eps_1}^{\eps_2} dx d\eps P(0,x,\eps) =
1$. \footnote{$P(0,\cdot,\cdot)$ is a density function at time zero over
  the opinion space and the interval of bounds of confidence,
  $P(0,x,\eps)dxd\eps$ represents the proportion of agents which hold
  opinions in $[x,x+dx]$ and bounds of confidence in
  $[\eps,\eps+d\eps]$.} For abbreviation we define the aggregated density
as $P(t,x) = \int_{\eps_1}^{\eps_2}d\eps P(t,x,\eps)$. With this we define the differential equation
\begin{align}\nonumber
\frac{\partial}{\partial t} P(t,x,\eps) =  \int\limits_S dy \left[ \left(\frac{}{}\right.\int\limits_{\eps_1}^{\eps_2}\right. &d\bar\eps  P(t,y,\bar\eps) \delta(x-M_1(y,P(t,\cdot),\bar\eps))
\left.\frac{}{}\right) \\ \label{eq:HKmaster1}
 & \left. - \quad P(t,x,\eps)\delta(y - M_1(x,P(t,\cdot),\eps))\frac{}{}\right]
\end{align}
The continuous-time density-based HK process is then defined as the solution of the initial value problem.
The discrete-time density-based HK process is the sequence $(P(t,\cdot,\cdot))_{t\in\N}$ recursively defined by 
\[
P(t+1,x,\eps) = P(t,x,\eps) + \Delta P(t,x,\eps)
\] 
with $\Delta P(t,x,\eps) = \frac{\partial}{\partial t} P(t,x,\eps)$.
\end{definition}

As for the density-based DW model it is important to distinguish the aggregated density $P(t,x)$ defined above and the density for a given bound of confidence $P(t,x,\eps)$ to understand how an agent looks at all agents regardless of their bound of confidence and takes into account all within her bound of confidence. 

For a homogeneous bound of confidence $\eps = [\eps_1,\eps_2]$ the dynamical Equation \eqref{eq:HKmaster1} reduces to
\begin{align}
\frac{\partial}{\partial t}P(t,x) =
\int_S\underbrace{\delta(M_1(y,P(t,\cdot),\eps) - x)P(t,y)}_{\textrm{fraction joining state $x$}} -
\underbrace{\delta(M_1(x,P(t,\cdot),\eps) - y)P(t,x)}_{\textrm{fraction leaving state $x$}}dy.  \label{eq:HKmaster2}
\end{align}

This can be transformed to the form
\begin{align}
\frac{\partial}{\partial t}P(t,x) = \left(\int\limits_{\{y \textrm{ is root of } x - M_1(y,P(t,\cdot),\eps)\}} dy\,\delta(y)\frac{P(t,y)}{\abs{M'_1(y,P(t,\cdot),\eps) }}\right) - P(t,x) \label{eq:HKmaster3}
\end{align}
with the $M'_1$ being the derivative with respect to $y$.
The function $M_1(y,P(t,\cdot),\eps)$ is monotone in $y$ for all $P$. Therefore $x - M_1(y,P(t,\cdot),\eps)$ has usually only one or no root. In the case of one root the integral in Equation \eqref{eq:HKmaster3} reduces to $\frac{P(y_0)}{\abs{M'_1(y_0,P(t,\cdot),\eps)}}$ with $y_0$ being that root. But there might be a continuum of roots due to a plateau of $x - M_1(y,P(t,\cdot),\eps)$ at zero, because $M_1(y,P(t,\cdot),\eps)$ is not strongly monotone.

It is shown in \cite{Fortunato2005b} that Equation \eqref{eq:HKmaster2}
conserves the mass. Further on it is claimed that the mean opinion is
also conserved. Although the derivation in \cite{Fortunato2005b} relies
on the symmetry of the initial opinion density around its mean the
authors forgot to mention that the mean opinion is not generally conserved for other initial conditions.

In general the right hand side of Equation \eqref{eq:HKmaster2} does not
depend continuously on $P(t,\cdot)$. (See example above.) But
unfortunately a $\delta$-peak in the density may evolve in this model by
computing the right hand side of Equation \eqref{eq:HKmaster2} even with
a given $P(t,\cdot)$ which does not contain a
$\delta$-peak. \footnote{Consider $P(t,\cdot)$ to be uniformly
  distributed on $[0,\frac{1}{3}] \cup [\frac{2}{3},1]$ and
  $\eps=0.5$. Then all the 'agents' with opinions in the interval
  $[0,\frac{1}{6}]$ have $\eps$-local mean equal to $\frac{1}{6}$ at that
  position we will have a $\delta$-peak.} Therefore the existence of
unique solutions to \eqref{eq:HKmaster2} is not generally assured for
every initial condition. Nevertheless they were computed in \cite{Fortunato2005b} with a fourth order Runge-Kutta algorithm and smooth looking dynamics were reached. 

But the picture changes drastically when we switch to the discrete-time
density-based HK model, which has been studied in
\cite{Lorenz2005a,Lorenz2006,LorenzPhD2007} as an interactive Markov
chain with 1000 and 1001 opinion classes. This coincides with the
discretization of the interval $[0,1]$ into 1000 bins in
\cite{Fortunato2005b}. In the discrete-time case two phenomena occur
which were not reported in \cite{Fortunato2005b}: Consensus striking back
for lower $\eps$ after a phase of polarization \cite{Lorenz2006} and the
importance of an odd versus an even number of opinion classes
\cite{Lorenz2005a}. Under an odd number, consensus is much easier to reach
because under an even number, the central mass contracts into two bins
which might be divided more easily compared to the odd case. In the odd case
the mass is stored in one bin, and mass in one bin can only move jointly
by definition of the process. We discuss the differences of continuous-time and discrete-time results in the next section. 

A numerical issue which is not clearly defined in
\cite{Fortunato2005b} is how moving mass is assigned to bins when
computing the right hand side of Equation \eqref{eq:HKmaster2}. Usually,
the $\eps$-local mean of a given bin lies not directly in a bin but
between two bins. So, there has to be a rule how to distribute it between
these two bins. In \cite{Lorenz2005a,Lorenz2006,LorenzPhD2007} it is done
proportionally to the distances from each bin. If the $\eps$-local mean
lies closer to one bin, this bin receives more of the moving mass. 

In this difference of continuous versus discrete time, the HK model
distinguishes from the DW model. Nevertheless, the possible fixed points of the
dynamics are the same as in the DW model. It is again every distribution
like $P(\infty,x) = \sum_{i=1}^r m_i\delta(x-x_i)$ with $|x_i-x_j|>\eps$
for all $i\neq j$ and $\sum_{i=1}^r m_i = 1$. This coincides with the
result for the agent-based model. But for both models it is not easy to
determine the limit density out of the initial density. Nearly every
study until now has studied only uniformly distributed initial densities. 

Example trajectories of the process can be found in \cite{Fortunato2005b,LorenzPhD2007}.

\section{Bifurcation Diagrams}

In this section we give the basic bifurcation diagrams for both homogeneous density-based models. A bifurcation diagram shows the location of clusters in the limit density versus the continuum of values of the bound of confidence $\eps$. So, one can determine the attractive cluster patterns for each bound of confidence and observe transitions of attractive patterns at critical values of $\eps$. The bifurcation diagrams serve as references for discussions on robustness to extensions of the homogeneous models and the impact of these extensions, e.g. on the critical bounds of confidence. 

\begin{figure}
	\centering
		\includegraphics[width=\textwidth]{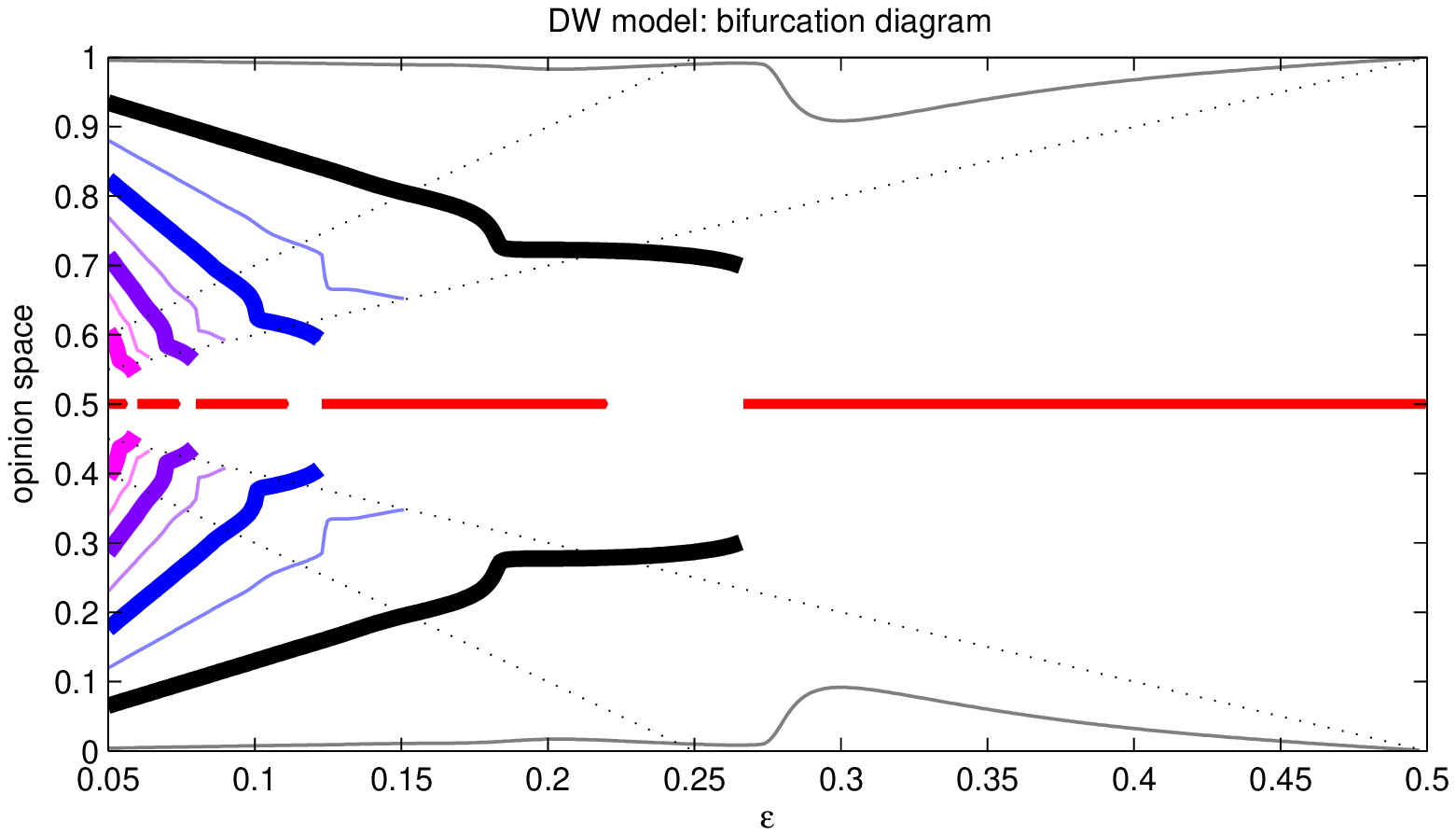}
		\includegraphics[width=\textwidth]{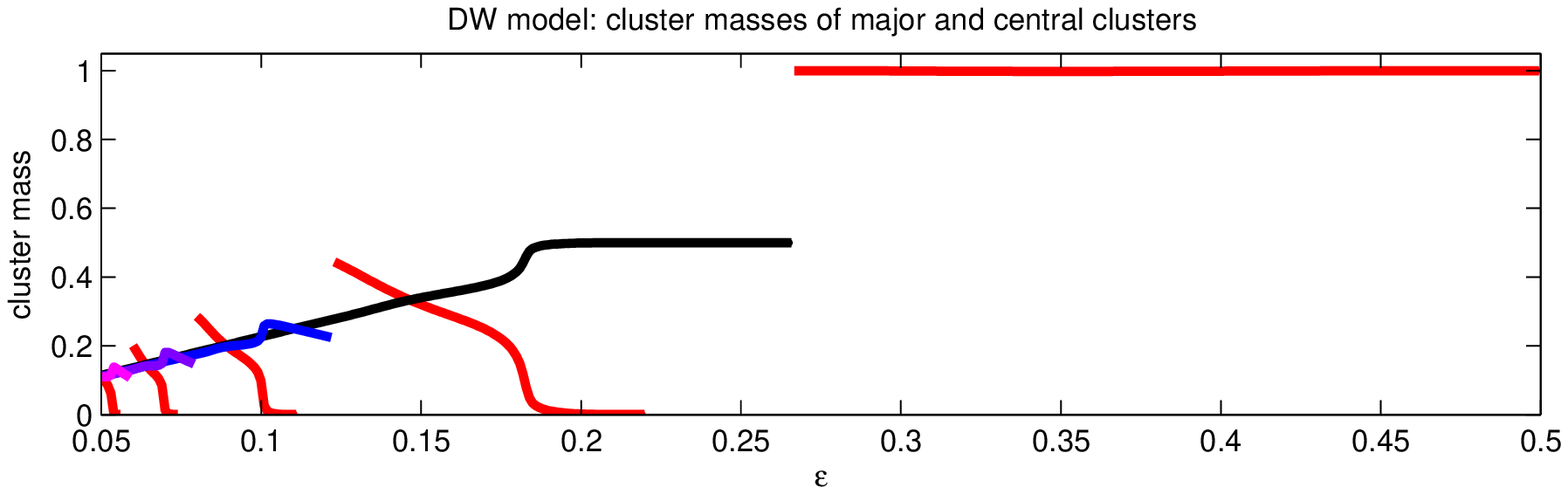}
		\includegraphics[width=\textwidth]{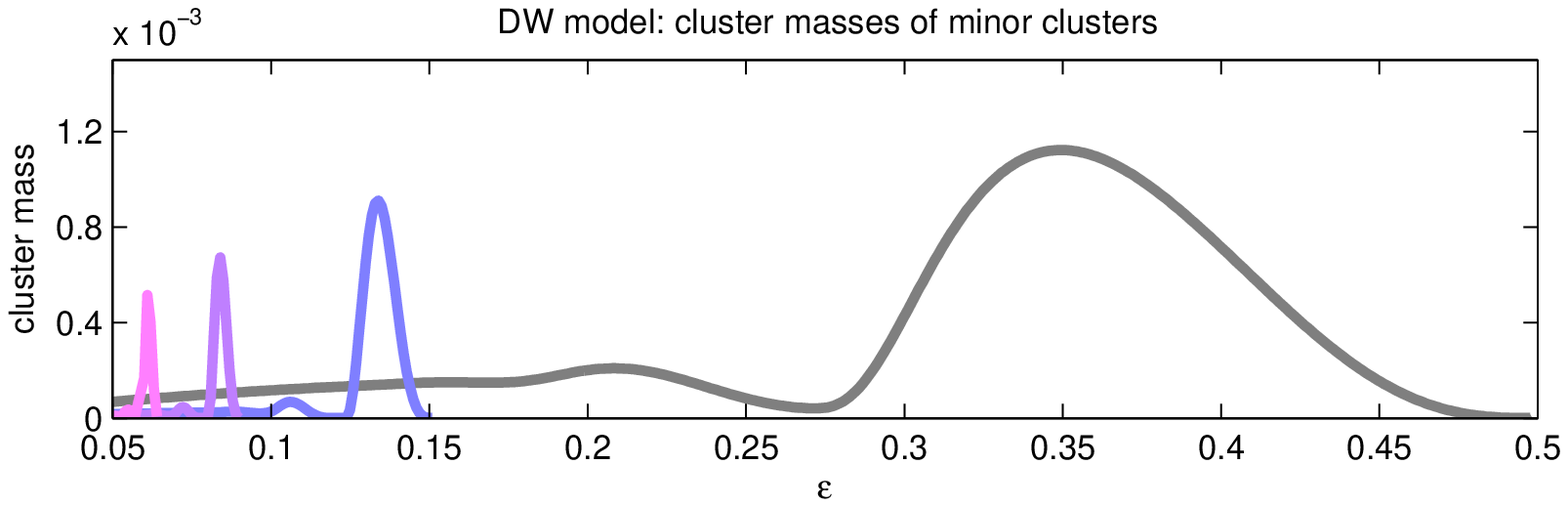}
	\caption{Bifurcation diagram for the DW model}
	\label{fig:bifDWconvGray}
\end{figure}

The top plot of Figure \ref{fig:bifDWconvGray} shows the bifurcation diagram for the homogeneous density-based DW model with uniform initial density in the opinion space $[0,1]$. 

A bifurcation diagram is made by computing the trajectory of the corresponding interactive Markov chain with 1001 opinion classes until clusters have evolved which spread not more than $\eps$ and are further away from each other than $\eps$. Clusters are determined by collecting intervals in the opinion space where the density is positive. Therefore, one has to decide on a threshold for which mass is regarded as zero. This threshold was $10^{-9}$ here. (For computational details see \cite{LorenzPhD2007}.)

The bifurcation diagram shown here resembles the bifurcation diagram shown in \cite{Ben-Naim2003}
but with a transformation of variables as $\Delta = \frac{1}{2\eps}$. In
\cite{Ben-Naim2003} $\Delta$ is the changing variable and determines the
opinion space $[-\Delta,\Delta]$ and the bound of confidence is fixed to
one. Here we have a fixed opinion space and a changing bound of
confidence because it is always done that way in agent-based simulations and
this framework can be extended to models with heterogeneous bounds of confidence. 

The fat lines in the top plot display the locations of clusters which
have a major mass, the thin lines the location of clusters with a minor
mass and the medium line the central cluster. Dotted lines are just for
orientation, they show the interval $[0.5-\eps,0.5+\eps]$. The two lower
plots show the masses of the related clusters. The most striking feature
is the existence of minor clusters at the extremes and between major
clusters. In agent-based simulations these minor clusters often do not
evolve because of low population sizes, nevertheless outliers are
frequently reported. These outliers make it difficult to count the number
of major clusters correctly. In \cite{Deffuant2000,Weisbuch2002} authors therefore decided to not count clusters with only one agent. The bifurcation diagram in \cite{Ben-Naim2003} clarified that minor clusters exist for structural reasons.

For $\eps\geq 0.5$ only one big central cluster evolves. As $\eps$
decreases bifurcations and nucleations of clusters occur. First, the
nucleation of two minor extremal clusters, then the bifurcation of the
central cluster into two major clusters, and third the rebirth of the
central cluster. For details see \cite{LorenzPhD2007,Ben-Naim2003}. This
bifurcation pattern is then repeated in shorter $\eps$-intervals. The length
of these intervals seems to scale with $\frac{1}{\eps}$. This can be
better seen in the bifurcation diagram of \cite{Ben-Naim2003}, where the
bifurcation pattern seems to repeat itself on intervals that converge
towards a length of about $2.155$. But this constant has only been derived numerically and
it is not clear that the bifurcation pattern will repeat that
regularly, although it looks very much like that. This result resembles the
rough $\frac{1}{2\eps}$-rule reported in \cite{Deffuant2000,Weisbuch2002} for
agent-based simulation, which says that the number of major clusters
after cluster formation is roughly determined as the integer part of
$\frac{1}{2\eps}$. 

A certain $\eps$-phase of interest is $\eps\in[0.27,0.35]$. As $\eps$
decreases the mass of the minor cluster decreases, too. Thus the mass of the
central cluster must increase slightly until the central cluster
bifurcates. Thus, surprisingly the
central cluster has a larger mass very close to the critical $\eps$. For further minor
clusters the same intermediate low-mass region exist. It even looks as if
the minor cluster disappears for a short intermediate $\eps$-range. The question if these gaps exist has
also been posed in \cite{Ben-Naim2003}. 

\begin{figure}
	\centering
		\includegraphics[width=\textwidth]{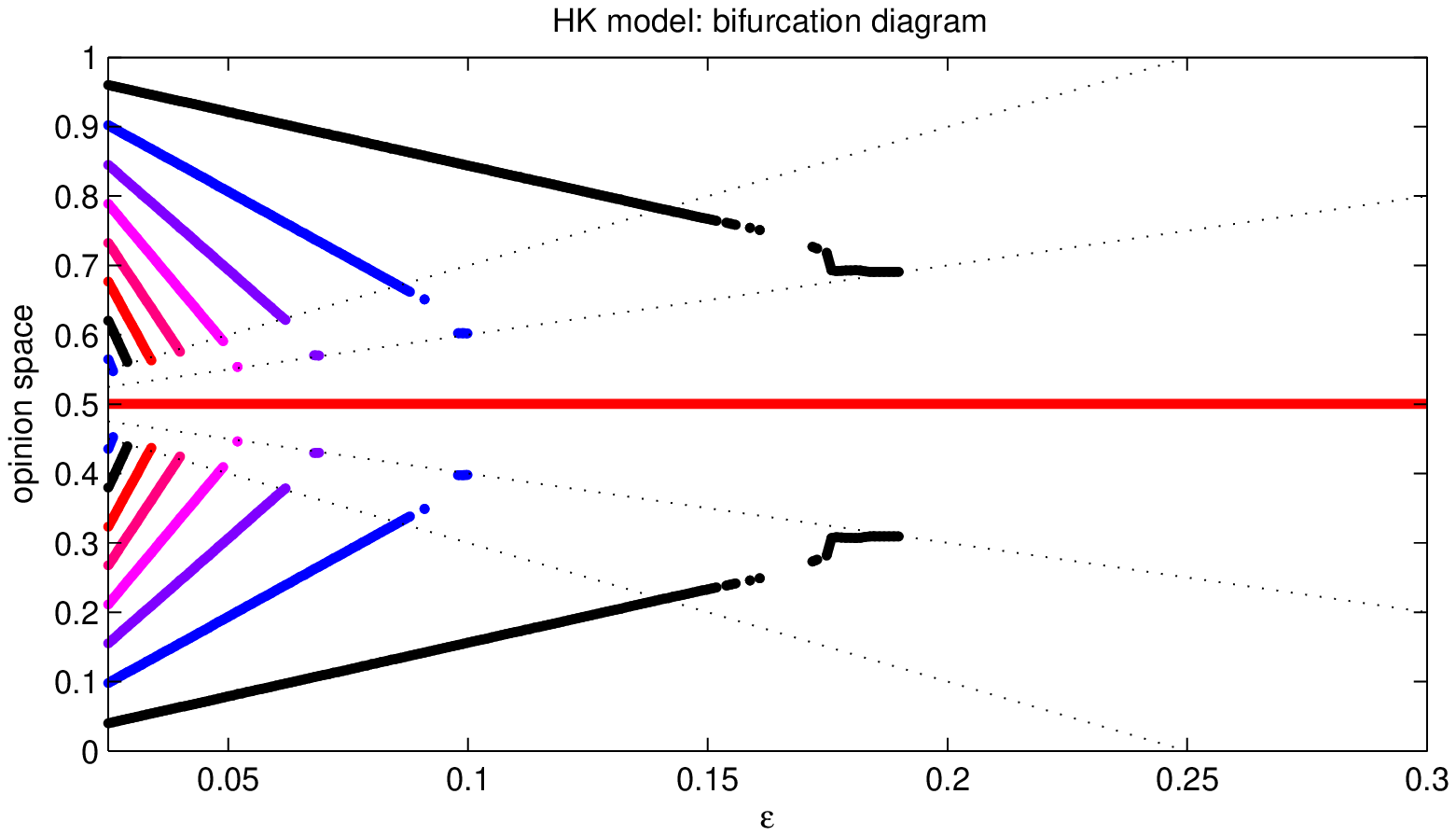}
		\includegraphics[width=\textwidth]{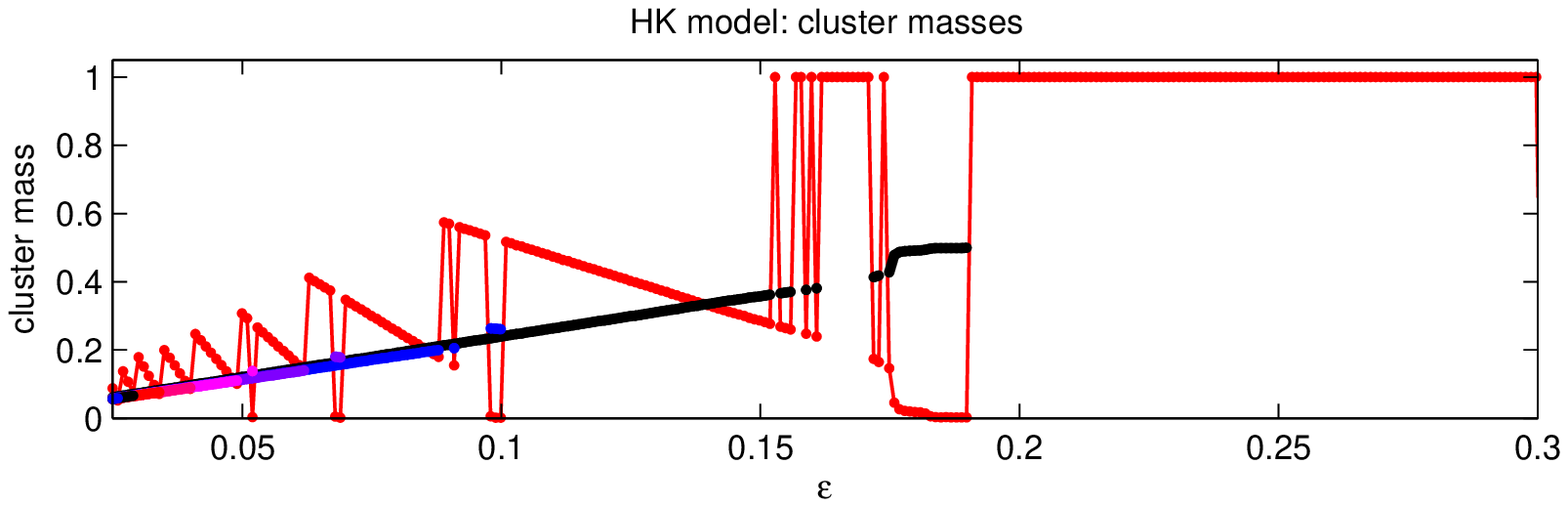}
		\includegraphics[width=\textwidth]{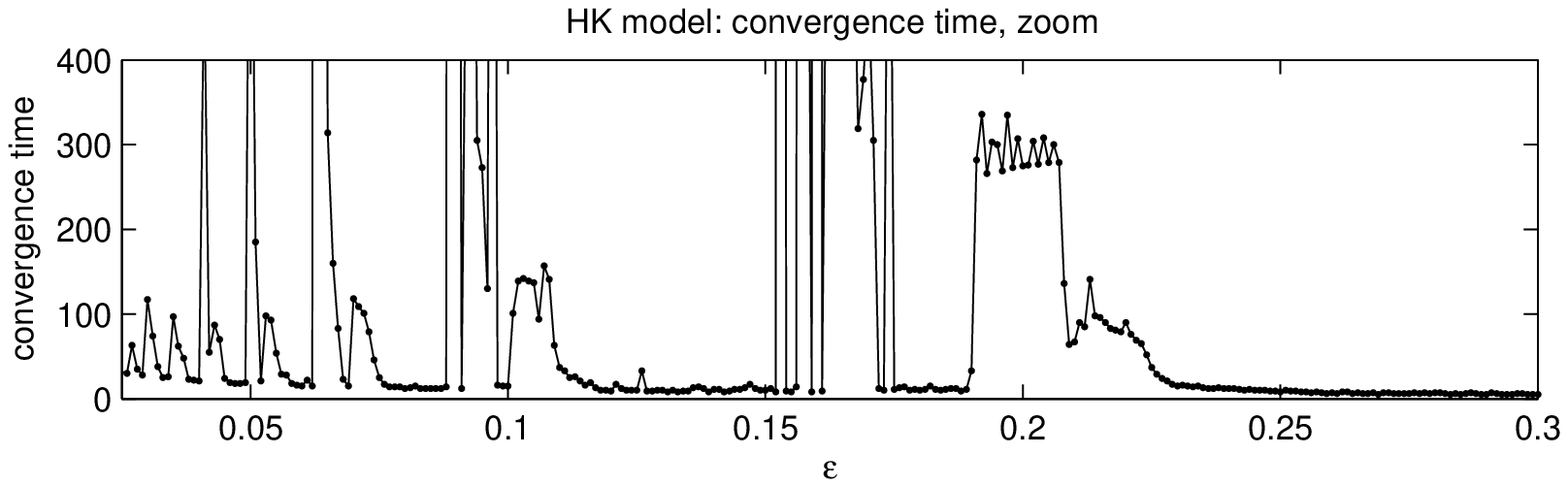}
	\caption{Bifurcation diagram for the HK model}
	\label{fig:bifHKconvGray}
\end{figure}

The top plot of Figure \ref{fig:bifHKconvGray} shows the bifurcation
diagram for the homogeneous density-based HK model with uniform initial
density in the opinion space $[0,1]$. The bifurcation diagram was computed the same way as for the DW
model. But the detection of the final density was much easier because the time-discrete process converges in finite time. (For computational details see \cite{LorenzPhD2007}.) 

The lines in the top plot display the locations of clusters. Unlike the
DW model the HK model has no minor clusters. The lower plot shows the masses of the related clusters. The convergence time (which is finite here) is the subject of the lowest plot. There are certain $\eps$-values were convergence time is really long. The plot shows a zoom, which does not display the longest times. The longest time was 21431.

For about $\eps\geq 0.19$ consensus is reached, but for values only
slightly above, it takes a long time. (See \cite{Fortunato2005b} for a
discussion why or \cite{LorenzPhD2007} for explanatory figures of
the processes.) So, $\eps=0.19$ is the consensus threshold. In
\cite{Fortunato2005b} the consensus threshold lies at about $\eps
=0.22$. This difference might have two reasons. First, it could be caused
by the differences of the continuous-time and the discrete-time process, and second it could result from the fact that they choose 1000 bins, so an even number. The computation in Figure \ref{fig:bifHKconvGray} was done with 1001 bins. This may really make a difference as pointed out in \cite{Lorenz2005a}. 

A striking feature of the bifurcation diagram is that after an
$\eps$-phase of polarization into two clusters (and a tiny central
cluster) there is a phase of even lower $\eps$ where we reach consensus
again at the expense of very long convergence time. Details about this phenomenon can be found in \cite{Lorenz2006}. Again this feature does not evolve in the bifurcation diagram of \cite{Fortunato2005b}. This might have again two reasons. First, the continuous-time approach, and second their termination criteria was so rough that it neglects the small movements which bring the two intermediate clusters together. Indeed there is a small hint for this reason in the bifurcation diagram in \cite{Fortunato2005b}: There is a small hill in the convergence time at the region where convergence to consensus happens in Figure \ref{fig:bifHKconvGray}.

So, the HK bifurcation diagram differs a lot from the DW bifurcation
diagram. Nevertheless, they are similar in the fact that the bifurcation pattern seems to repeat on time scales scaling with $\frac{1}{\eps}$. Again, there is no proof of this universal scaling. Moreover there is evidence that the regularity gets destroyed for lower $\eps$. (See \cite{Lorenz2006} for details, the evidence can be seen by studying onesided dynamics.)

\begin{figure}
	\centering
		\includegraphics[width=\textwidth]{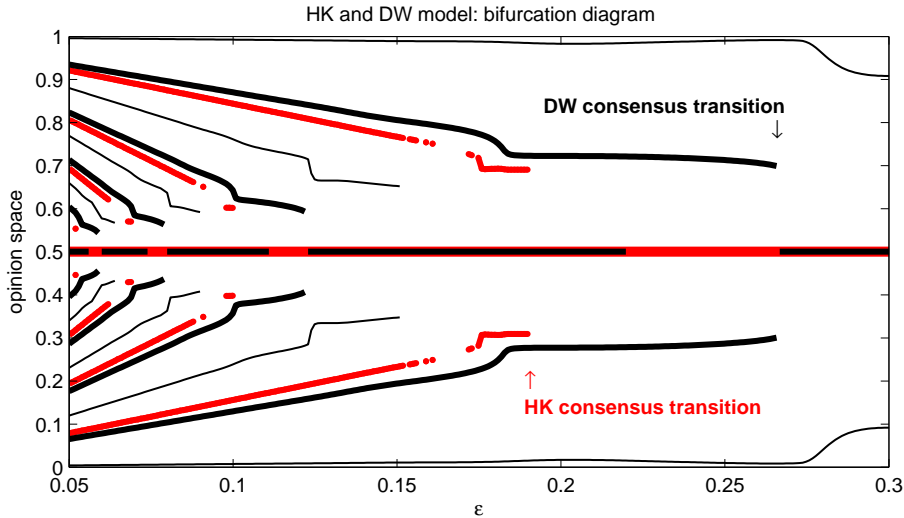}
	\caption{Bifurcation diagram for the DW and HK model}
	\label{fig:bifHKDW}
\end{figure}

Figure \ref{fig:bifHKDW} shows both bifurcation diagrams in one plot to
compare the differences. (Black lines are DW model, red (color online)
lines ar HK model.) Notice the different critical values of $\eps$ for the consensus transition. The consensus transition here is the bifurcation point between two major clusters and a big central cluster. So, we neglect the minor clusters in the DW model for this definition. 

The DW and HK model have been studied with several extensions, which we will review in the following with respect to the bifurcation diagrams presented here.

\section{Extensions}

The Definitions 1, 2, 3 and 4 already contain the possibility of multidimensional opinions and heterogeneous bounds of confidence for agents, respectively subpopulations. Several further extensions which are based on continuous opinions and bounded confidence haven been proposed and analyzed to some extent. A good example of the overwhelming number of ideas which need to be checked is \cite{Hegselmann2004}. We will review the known results and propose some open questions. 

\subparagraph{Nonuniform initial conditions}

This topic is not really an extension but a very important issue. Nearly
every study used random and uniformly distributed initial opinion
profiles or initial densities which are uniformly distributed in the
opinion space. The right classification of initial profiles leading to
structurally different final cluster configurations remains a task for
future work. At the agregate level the agent-based simulations are in fair agreement with the density-based processes, although random fluctuations play important roles in individual trajectories even for large populations. This topic is also discussed in \cite{Jacobmeier2006}.

\subparagraph{Multidimensional opinions}

Until now, we have only considered the opinion space $[0,1]$. Other lengths
of intervals will not deliver new insight, because they will give the
same structural behavior only on scaled $\eps$-levels. So, we have a
complete picture in studying the unit interval. This changes when
investigating multidimensional opinions. The usual first step is to study
$S=[0,1]^2$, see \cite{Fortunato2005b, Lorenz2003b, Pluchino2006}. Final
clusters then form a regular square latice but with some interferences in
the center, where a central cluster may or may not exist. See
\cite{Pluchino2006} for a circular disk as the opinion space in an agent-based HK model and how it can have a positive impact on finding consensus by preventing corner clusters. See \cite{Lorenz2003b} for a hexagonal and a triangular opinion spaces. 

Another interesting parameter, which plays a role in more dimensions is the
distances measure which is a norm of the difference of two opinions in
this paper, but of course any metric might be a candidate. The most common $p$-norms ($\norm{x}_p = (\sum_{i=1}^d \abs{x_i}^p)^\frac{1}{p}$) have been studied. These are $p=1,2,\infty$ in \cite{Lorenz2003b} for the agent-based HK model (see also \cite{Pluchino2006}) and $p=2,\infty$ in \cite{Fortunato2005b} for the density-based HK model. There is some impact of the norm but the interplay of the distance measure and the shape of the opinion space is not well understood.

There are only two studies about the impact of the dimensionality parameter $d$ itself. In \cite{Lorenz2006} $d$-dimensional unit simplexes ($S=\triangle^d=\{x\in\R^{d+1}_{\geq 0} \,|\, \sum_{i=1}^{d+1}x_i=1\} $) as opinion spaces were studied up to $d=7$. The simplex opinion space represents opinions which are under budget constraints. One opinion is a distribution of resources to $d+1$ projects (the opinion issues) but with a fixed amount of resources. It has been shown that raising $d$ lowers the consensus threshold in the agent-based DW as well as in the agent-based HK model. But the consensus threshold cannot be pushed down to zero. So, there is a limit above which raising the number of issues $d+1$ brings only marginal positive effects on fostering the chances for consensus.

The budget constraints are crucial for this positive effect as it is shown in \cite{Lorenz2007Helb}. If opinions are not under budget constraints like in $S=[0,1]^d$ raising $d$ raises the consensus threshold. This has been checked for DW and HK communication, $1$- and $\infty$-norm. 

In a nutshell, if you want to foster consensus, bring more interrelated
issues into the discussion. If you want to prevent consensus, bring more
independent issues into the discussion.

One source of this effect is the growth of the volumes of opinion spaces
compared with the growth of volumes of areas of confidence with rising
dimension. Under budget constraints the volume of the area of confidence
growth faster than the volume of the opinion space. It is the other way
round without budget constraints. 

Another interesting new approach in multidimensional continuous opinion dynamics is to study dynamics of structured attitudes and opinions \cite{Urbig2005}, where attitudes are multidimensional but opinions are an aggregation of attitudes and dynamics happen on 
both levels.

\subparagraph{Heterogeneous bounds of confidence}

The first brief study of heterogeneous bounds of confidence has been done in \cite{Weisbuch2002} for the DW model with a population of agents with two levels of bounds of confidence, the open-minded and the closed minded agents. They studied a set of 192 closed-minded agents ($\eps_1 = 0.2$) and eight open-minded agents ($\eps_2 = 0.4$) and saw that in the short run, the cluster pattern of the closed-minded (two big clusters) dominated while in the long run, the one of the open-minded (consensus) evolved. 

Meanwhile it has been shown by analyzing the density-based DW model
\cite{LorenzPhD2007} that there is even more impact of
heterogeneity. Consensus can be achieved by mixing closed- and
open-minded agents even if both bounds of confidence are far below the
critical value of the consensus transition (e.g. $\eps_1 = 0.11$ and
$\eps_2=0.22$). But on the other hand drifting of clusters of open-minded towards clusters of closed-minded agents becomes a generic feature of dynamics which can amplify some asymmetric disturbances in the initial profile. So in the end a final consensual cluster may lie very far from the initial average opinion. 

This drifting phenomenon is the source of the drift to the extremes studied in \cite{Deffuant2002,Weisbuch2005}. There, very few closed-minded agents are initially set to the extremes of the opinion space and open-minded agents are set in between. So, very stable clusters of extremists always evolve at the extremes and the question is where the central agents go. They may form a central cluster, divide to both extremes or drift together to one extreme, depending on the threshold of the open-minded agents.

The HK model shows similar drifting phenomena when heterogeneous bounds of confidence are introduced. Further on, another phenomenon is `sitting between the chairs' of the open-minded agents. It might happen that two clusters of closed-minded agents form and open-minded agents form a cluster in between although they take the opinions of both closed-minded clusters into account. 

\subparagraph{Social networks and communication regimes}

Another issue which has been studied is the influence of social
networks. Communication is not only restricted by the bounded confidence
assumption but also by a given static social network. In
\cite{Fortunato2004a} and \cite{Fortunato2005} the agent-based DW and HK
model has been set up on networks with scale-free link distribution \`{a}
la Barab\'{a}si-Albert, random graphs \`{a} la Erd\"{o}s-R\'{e}nyi, and
square lattices. They gave evidence for the fact that in the DW model for
large enough networks consensus is always reached for $\eps>0.5$, which
was later proven in \cite{Lorenz2007}. Consensus here means full
consensus, even without small extremist clusters. Further on, there is
evidence that the consensus threshold for the HK model only remains to be
(very roughly) about $0.2$ if the average number of links scales with the
number of agents if these numbers go to infinity. Otherwise it switches to $0.5$. In \cite{Stauffer2004} the DW model was again set on a Barab\'{a}si-Albert graph and gives evidence that the number of final clusters scales now with the number of agents and not only with $\frac{1}{\eps}$. (See also \cite {Stauffer2004a} and \cite{Weisbuch2004a} for results and some different versions of DW dynamics on scale-free networks.)
In \cite{Jacobmeier2004} a combination of multidimensional opinion dynamics on scale-free networks is done. 

The extremism version of the DW model proposed in \cite{Deffuant2002} has been set on small world graphs \`{a} la Duncan-Watts leading to the fact that the drift to one extreme only appears beyond a critical level of random rewiring of the regular network \cite{Amblard2004}.

In \cite{Urbig2004} a general model has been proposed which includes the DW and the HK model as special cases. A fixed number of $m$ agents meet at each time step and perform a HK step. There is evidence that the transition from the HK to the DW model is somehow smooth. 

Finally, in \cite{Lorenz2007} it is emphasized that only a static network may modify the dynamics but also the choice of random communication partners in the agent-based DW model may be important. Such a choice is called communication regime. It is shown that enforcing as well as preventing consensus is possible for a for huge $\eps$-phase if it is possible to manipulate who meets whom at what time. 

\subparagraph{Convergence parameter, cautiousness or quality}

The original agent-based DW model in \cite{Deffuant2000} contained a
convergence parameter $\mu \in ]0,0.5]$ which controls how far an agent
moves to the opinion of the other if they are closer than $\eps$ to each
other. The model presented here is the case $\mu = 0.5$. For lower $\mu$
agents make shorter jumps towards the other opinion. Therefore, it can
also be called cautiousness parameter. In a brief analysis of the basic
model this parameter has been shown to affect only convergence time. Low
$\mu$ leads to longer convergence time. But in \cite{Laguna2004} it is
shown by simulation that lower $\mu$ also reduces the sizes of the minor
clusters. So more cautious agents are not that likely to produce minor
clusters. Further on, $\mu$ can have an interplay with other parameters
when they are introduced, e.g. the number of agents which meet in a time
step in \cite{Urbig2004}. It is possible to include the cautiousness
parameter in a density based version of the model \cite{Lorenz2007}. In
\cite{Assmann2004} there is a study of the DW model on a growing
scale-free network with quality differences of opinions. High quality is
modeled as low $\mu$, which corresponds to high cautiousness. But quality is assigned to an opinion not to an agent. So, an agent's cautiousness changes when she changes her opinion.

\subparagraph{Miscellaneous}

In \cite{Hegselmann2002} the agent-based HK model is analyzed for biased
confidence intervals. So, for instance agents look more to the left than
to the right. This can lead to consensus at more extreme positions. 

The agent-based DW model has been extended two times to smooth bounds for the confidence interval, called the relative agreement model \cite{Deffuant2002} and smooth bounded confidence model \cite{Deffuant2004}. The idea is that the attraction of agents should be lower if they are far from each other but still not totally zero. Results remained quite similar. Structural effects remain to be discovered.

Another extension is the introduction of repulsive forces when agents are
far away from each other. This is studied in the DW setting in
\cite{Jager2005} motivated by social judgment theory. There were
absorbing boundaries at the extremes and in the center dynamics were
similar to the usual clustering.

The agent-based HK model has been extensively studied with other means
than the arithmetic mean in \cite{Hegselmann2004a} leading to very
similar clustering behavior except for what is called a random
mean. But positions of clusters change a lot in comparison to the original model.  Proofs of covergence are also possible for abstract means.

Communication strategies of agents in the agent-based DW model were a topic in \cite{Lorenz2007}. Agents may be balancing and try to find communication partners from different directions or they may be curious and try to find communication partners of the same direction after a successfull communication. Both lead to better chances for consensus. But there is an interplay with the cautiousness parameter $\mu$. For balancing agents it is better to be cautious to enhance chances for consensus, but for curious agents it is not good to be cautious.

Urbig \cite{Urbig2003} argues that it is important to distinguish opinions and attitudes (e.g. about a product). The attitude is a continuous variable which each agent holds internally. The attitude becomes an opinion when verbalized and communicated. With this process it becomes discrete. In this terms this paper is about `attitude dynamics'.

From a more philosophical point of view the topic of finding the truth is
incorporated in the HK model in \cite{Hegselmann2006a} as an agent who
automatically sticks to the `true' value. Questions are then, how many
agents who go for the truth do we need such that all agents end up finding the truth. In \cite{Malarz2006} the idea is used for the DW model. 

\section{Conclusions and open problems}

The models of continuous opinion dynamics under bounded confidence have
been shown to capture some phenomena which have are not covered by traditional binary opinion models. These phenomena are dynamic formation of different numbers of clusters with characteristic location and size patterns, and drifting of clusters when bounds of confidence are heterogeneous. 
Also from a modeling point of view continuous issues are different from discrete issues. Continuous opinions are more related to negotiation problems or fuzzy attitudes which do not really match with a yes or no decision.

The master equation approach has lead to the density-based models which
give better insight on attractive states of the agent-based models. The numerical integration of the master equation is something between a full analytical understanding and the analysis by Monte Carlo simulations. But on the other hand it is still important to match Monte Carlo runs with computations of densities because we are not only interested in dynamics in the thermodynamic limit but especially in dynamics of finite populations (of various sizes). See also \cite{ToralEtal2006} for results when finite population sizes are essential for opinion dynamics. Finite size effects play an important role even in very large populations (especially in the HK model) but on the other hand the density-based models delivers attractive final cluster patterns for the agent-based model even for very low numbers of agents. But as ususal for lower population sizes a Monte Carlo run does not always match the attractive cluster pattern predicted by the density-based model. A quantification to what extend the density-based model matches with the agent-based models of a given population size would be a great improvement.

From an analytical point of view the most important remaining issue is
the understanding of the impact of the initial profile or initial density
on the final outcome. The task is to classify initial profiles which lead to essentially different cluster patterns. The next important problem is to understand the drifting phenomena under heterogeneous bounds of confidence. Here, the master equation outlined above seems to be rewarding.
But still the universal scale for bifurcation points is not proven today.
Even the convergence to a stable cluster formation itself is
not formally proven for the density-based models and the agent-based
models under heterogeneous bounds of confidence. Trying to prove all that
should also lead to more insight on dynamics.  

From a theoretical modeling point of view, it would be interesting to
classify all models of continuous opinion dynamics which show
stabilization of opinions in clusters. 

From an application point of view, a deeper analysis of continuous opinion dynamics may lead to a better design of decision and discussion processes in large populations. Fostering the chances of finding consensus is a good starting point for research about extensions. The next level is then to foster consensus within the bounds of not too large convergence times.

\end{document}